\documentclass[twocolumn]{aastex701}
 
\usepackage{amsmath}
\usepackage{amssymb}

\begin{document}

\title{COSMOS-Web: Galaxy Size and Surface Brightness Evolution at Rest-Frame 1.22 $\mu$m Since $z=3$}

\author[orcid=0000-0002-3462-4175]{Si-Yue Yu}
\affiliation{Department of Astronomy, Xiamen University, Xiamen, Fujian 361005, People's Republic of China}
\email[show]{syu@xmu.edu.cn}

\author[orcid=0000-0002-9382-9832]{Andreas L. Faisst}
\affiliation{Caltech/IPAC, 1200 E. California Blvd., Pasadena, CA 91125, USA}
\email{afaisst@caltech.edu}

\author[orcid=0000-0002-2853-3808]{Taotao Fang}
\affiliation{Department of Astronomy, Xiamen University, Xiamen, Fujian 361005, People's Republic of China}
\email{fangt@xmu.edu.cn}

\author[orcid=0009-0005-3133-1157]{Greta Toni}
\affiliation{Dipartimento di Fisica e Astronomia ``Augusto Righi'', Alma Mater Studiorum Università di Bologna, via Gobetti 93/2, I-40129 Bologna, Italy}
\affiliation{INAF-Osservatorio di Astrofisica e Scienza dello Spazio di Bologna, via Gobetti 93/3, 40129 Bologna, Italy}
\affiliation{Zentrum f\"{u}r Astronomie, Universit\"{a}t Heidelberg, Philosophenweg 12, D-69120, Heidelberg, Germany}
\email{greta.toni4@unibo.it}

\author[orcid=0000-0002-3473-6716]{Lauro Moscardini}
\affiliation{Dipartimento di Fisica e Astronomia ``Augusto Righi'', Alma Mater Studiorum Università di Bologna, via Gobetti 93/2, I-40129 Bologna, Italy}
\affiliation{INAF-Osservatorio di Astrofisica e Scienza dello Spazio di Bologna, via Gobetti 93/3, 40129 Bologna, Italy}
\affiliation{INFN-Sezione di Bologna, Viale Berti Pichat 6/2, I-40127 Bologna, Italy}
\email{lauro.moscardini@unibo.it}

\author[0000-0002-3560-8599]{Maximilien Franco}
\affiliation{Université Paris-Saclay, Université Paris Cité, CEA, CNRS, AIM, 91191 Gif-sur-Yvette, France}
\email{maximilien.franco@cea.fr}

\author[orcid=0000-0003-2716-8332]{Rasha M. Samir}
\affiliation{Department of Astronomy, National Research Institute of Astronomy and Geophysics (NRIAG), Cairo, Egypt}
\email{rasha.samir@nriag.sci.eg}

\author[0000-0002-3301-3321]{Michaela Hirschmann}
\affiliation{Institute of Physics, GalSpec, Ecole Polytechnique Federale de Lausanne, Observatoire de Sauverny, Chemin Pegasi 51, 1290 Versoix, Switzerland}
\affiliation{INAF, Astronomical Observatory of Trieste, Via Tiepolo 11, 34131 Trieste, Italy}
\email{michaela.hirschmann@epfl.ch}

\author[orcid=0000-0003-4832-9422]{Xiaoxia Zhang}
\affiliation{Department of Astronomy, Xiamen University, Xiamen, Fujian 361005, People's Republic of China}
\email{zhangxx@xmu.edu.cn}

\author[orcid=0009-0004-2523-4425]{Gavin Leroy}
\affiliation{Institute for Computational Cosmology, Department of Physics, Durham University, South Road, Durham DH1 3LE, United Kingdom}
\email{gavin.leroy@durham.ac.uk}

\begin{abstract}

We present the evolution of galaxy size and surface brightness in the rest-frame {\it J} band (1.22 $\mu$m), tracing the stellar mass distribution, over $0.5 \leq z \leq 3$, using a sample of 15,420 galaxies with stellar masses $M_\star=10^{10}$--$10^{11.5}\ M_{\odot}$ from the JWST COSMOS-Web survey. The rest-frame {\it J}-band effective radius ($R_{e,J}$) is obtained from previous measurements and mapped from the available JWST/NIRCam filters, while the surface brightness ($\mu_J$) is corrected for dust extinction and cosmological dimming. At a characteristic mass of $M_\star = 5 \times 10^{10}\ M_{\odot}$, star-forming galaxies exhibit a size evolution of $R_{e,J} \propto (1+z)^\beta$ with $\beta = -0.92 \pm 0.04$, falling between previously reported shallower and steeper measurements. Quiescent galaxies evolve more rapidly, with $\beta = -1.34 \pm 0.05$, consistent with earlier studies. Among star-forming galaxies, lower-mass systems ($10^{10}$ to $10^{10.5}\ M_{\odot}$) show slower ($\beta=-0.66\pm0.02$) size evolution compared to their higher-mass counterparts. Furthermore, the surface brightness brightens toward higher redshifts, scaling as $\mu_J \propto -2.5 \log(1+z)^\gamma$. We find $\gamma = 3.07 \pm 0.08$ for star-forming galaxies and $\gamma = 3.70 \pm 0.08$ for quiescent galaxies. We also find that massive star-forming galaxies ($M_\star > 10^{10.5}\ M_{\odot}$) exhibit similar $\mu_J$ values at fixed redshift, independent of mass. Finally, we demonstrate that the observed surface brightness evolution is driven by the combined evolution of galaxy luminosity and size.

\end{abstract}


\section{Introduction} \label{intro}

Galaxy size is a fundamental structural property that can be measured directly from imaging, and its evolution with redshift provides insight into how galaxies assemble and redistribute their stellar mass over cosmic time. In simple disk-formation models, galaxy size is linked to the virial scale of the host dark matter halo, implying that disks assembled at earlier epochs should be more compact \citep{Mo1998}. Subsequent evolution can further modify galaxy sizes through processes such as inside-out growth \citep[e.g.][]{Guo2011}, stellar migration \citep[e.g.][]{Debattista2006, Minchev2012, Yu2025}, clump-induced scattering \citep[e.g.][]{Bournaud2007, Yu2025}, and mergers \citep[e.g.][]{Hopkins2008}.

Extensive HST studies established that galaxy sizes at a given stellar mass decrease toward higher redshift, with quiescent galaxies evolving more rapidly than star-forming systems \citep[e.g.,][]{Bouwens2004, Daddi2005, Trujillo2007, Buitrago2008, Oesch2010, Mosleh2012, vanderWel2014, Faisst2017, Whitney2019, Yang2021}. For example, at fixed stellar mass, \citet{vanderWel2014} found that rest-optical sizes scale approximately as $(1+z)^{\beta}$ with $\beta=-0.75$ for star-forming galaxies and $\beta=-1.48$ for quiescent galaxies over $0\leq z\leq3$. 
In contrast, \citet{Faisst2017} show that in the ultra-massive regime ($M_\star > 10^{11.4},M_\odot$) at $z \lesssim 2$, the two populations may follow similar evolutionary trends.
However, HST observations at $z\sim3$ probe rest-frame wavelengths of $\sim0.4 \micron$ with the F160W filter, potentially introducing systematic biases in size measurements between star-forming and quiescent galaxies. The advent of JWST, with its unprecedented infrared sensitivity and resolution, has opened a new window for studying galaxy structure at high redshift. Nevertheless, recent JWST studies have yet to converge on a consistent picture, especially for star-forming galaxies.
\citet{Ward2024} reported $\beta=-0.63\pm0.07$ for rest-frame optical sizes, while \citet{Martorano2024} found a similar but slightly shallower slope, $\beta=-0.69\pm0.1$ at rest-frame $1.5$ \micron. At higher redshift, \citet{Allen2025} derived an intermediate value of $\beta=-0.807\pm0.026$ in the rest-frame optical. In contrast, \citet{Yang2025} reported a substantially steeper evolution, $\beta=-1.21\pm0.05$. Focusing on star-forming brightest group galaxies, \citet{Gozaliasl2025} obtained $\beta=-0.96\pm0.07$.
This dispersion suggests that the redshift evolution of galaxy size is still not fully settled, likely owing to differences in rest-frame wavelength and measurement methodology. A new independent measurement based on a large homogeneous sample and a dust-insensitive rest-frame band is therefore needed.

Galaxy size alone, however, does not fully describe how stellar light is distributed within galaxies. Surface brightness provides a complementary observable by combining luminosity and size, and its evolution reflects the interplay between structural growth and luminosity evolution \citep{Roche1998, Barden2005, Conselice2014}. Previous studies based on ground-based and HST imaging found that galaxies tend to have brighter intrinsic surface brightness at higher redshift, but the inferred trends remain uncertain because of limited spatial resolution, bandpass differences, and heterogeneous measurement methods \citep{Schade1995, Schade1996, Roche1998, Barden2005, Holden2005, Ichikawa2010, Whitney2020}. JWST now makes it possible to revisit this problem at rest-frame near-infrared wavelengths, where the effects of dust are reduced. A joint analysis of size and surface brightness in the rest-frame near-infrared can thus provide cleaner constraints on the structural and photometric evolution of galaxies.

In this work, we measure the evolution of galaxy size and surface brightness in the rest-frame {\it J} band (1.22 \micron) over $0.5 \leq z \leq 3$ using COSMOS-Web, 
the largest extragalactic survey conducted in JWST Cycle 1, covering 0.54 deg$^2$ \citep{Casey2023}. This dataset provides both the sample size and the image quality needed to re-examine galaxy size evolution and to place robust constraints on the corresponding evolution of surface brightness in a uniform framework. Throughout this paper, we adopt AB magnitudes, the \cite{Chabrier2003} initial mass function, and a flat $\Lambda$CDM cosmology with $(\Omega_{\rm m},\Omega_{\Lambda},h)=(0.27,0.73,0.70)$.

\section{Sample and Data} \label{sec:sample}

This study builds upon \citet{Yu2026}, which analyzed a parent sample of 17,754 galaxies from the JWST COSMOS-Web survey over photometric $0<z<3.5$ with stellar mass $M_*\geq10^{10}M_\odot$, including single-Sérsic fitting using {\tt IMFIT} \citep{Erwin2015}. Their sample excludes galaxies undergoing mergers or affected by close neighbors or bright stars. We further exclude \textbf{8} Little Red Dots identified by \cite{Akins2025} and \textbf{35} X-ray sources, likely AGNs, identified by \cite{Marchesi2016}. 
We further restrict the sample to $10^{10}M_\odot\leq M_*\leq 10^{11.5}M_\odot$ and redshift  $0.5 \leq z \leq 3$. The higher $z$ limit ensures coverage of the rest-frame {\it J} band, despite minor extrapolation, while the lower limit is set because of the low galaxy number due to small cosmic volume. These criteria yield a final sample of 15,420 galaxies. We acquire the best-fit effective radii ($R_e$) from S\'{e}rsic modeling at JWST/NIRCam F115W, F150W, F277W, and F444W band from \cite{Yu2026}.
The involved PSFs are hybrid PSFs, which combine empirical cores with {\tt WebbPSF}-modeled \citep{Perrin2014} outer profiles, ensuring accurate characterization of both central structures and extended wings across all bands (see \citealt{Yu2026} for details).

At high redshift, galaxies detected in F115W or F150W can be indistinguishable from, or only marginally above, the background. Following \citet{Yu2026}, we exclude bands in which the mean flux within the galaxy segmentation falls below the background level, and instead retain only those with sufficient signal-to-noise ratio. To derive rest-frame $J$-band effective radius ($R_{e,J}$), we linearly fit $\log(R_e/{\rm kpc})$ against logarithmic values of rest-frame \textbf{central} wavelength \textbf{of each filter} to determine the $1.2\ \micron$ value, and correct it using the residual between the observed and best-fit measurements from the closest NIRCam filter. Examples of the wavelength dependence of $R_e$ are provided in the Appendix.

We note that $R_e$ and the Sérsic index ($n$) are coupled parameters in Sérsic profile fitting, and their covariance can affect measured structural trends. However, a measured $R_e$--$n$ trend should not be interpreted as a purely artificial fitting degeneracy, as it also reflects underlying correlations among galaxy luminosity, effective radius, and effective surface brightness \citep{Trujillo2001}. More importantly, image simulations by \citet{Yu2023} showed that $R_e$ can be robustly recovered without significant systematic bias, even for small galaxies. Therefore, the covariance between $R_e$ and $n$ is unlikely to drive the size evolution measured in this work.

Moreover, the $R_e$ mainly traces the radial distribution of the inner light and can depend on the relative contribution of different structural components. Recent studies have proposed physically motivated size definitions based on the outermost radius where gas has efficiently collapsed and formed stars, or where the stellar mass density or surface brightness profile shows a truncation associated with a drop in past or ongoing in-situ star formation \citep{Trujillo2020, Chamba2022,Buitrago2024}. These measures can more directly trace the boundary of the main in-situ stellar component and the physical extent of disk growth. Thus, $R_e$ and these physically motivated size definitions provide complementary views of galaxy structure. Applying such definitions in future studies will be valuable for further investigating galaxy evolution.

Stellar masses ($M_*$), photometric redshifts ($z$), rest-frame absolute magnitudes in the NUV, {\it r}, and {\it J} bands, and color excess $E(B-V)$ are adopted from the COSMOS2025 catalog (v1.1)\footnote{\url{https://cosmos2025.iap.fr}} \citep{Shuntov2025Cat}. $M_*$ refers to the current stellar mass, rather than the total stellar mass ever formed over the history of the galaxy.

\begin{figure*}
  \centering
  \includegraphics[width=2\columnwidth]{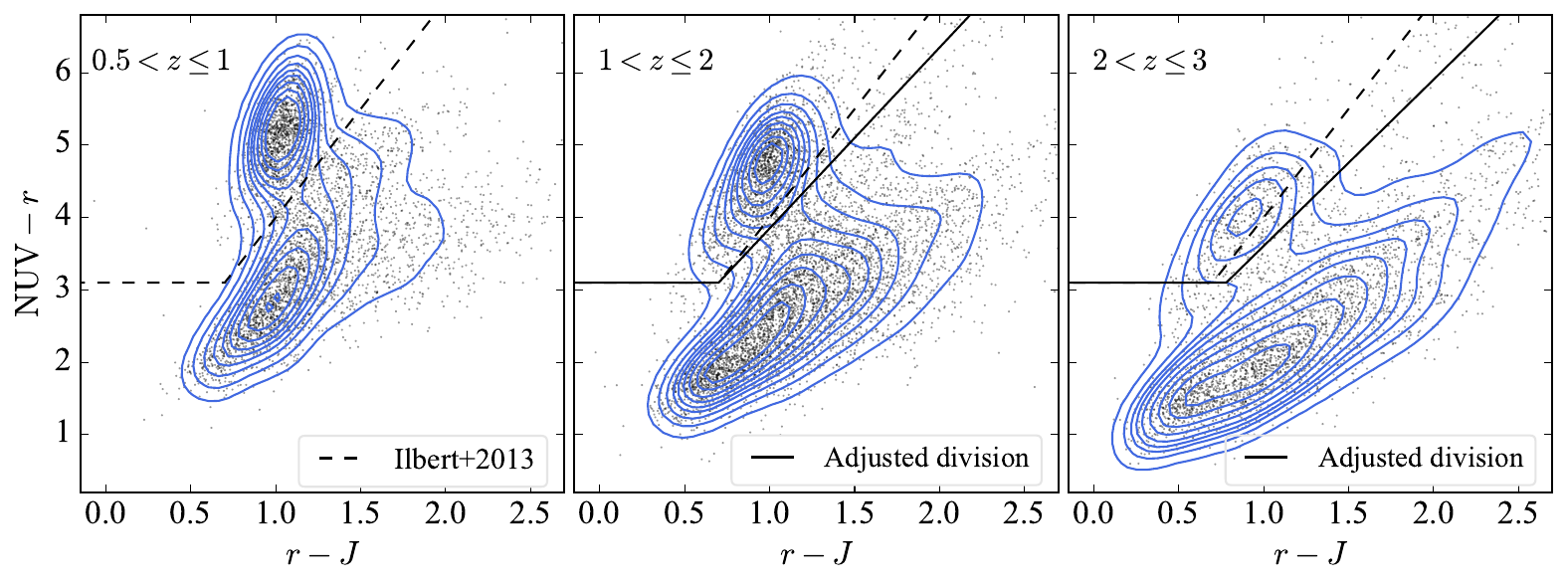}
  \caption{Rest-frame color-color diagram for our galaxy sample in three redshift bins. The individual measurement is marked by the background gray dot. Contours indicate regions enclosing a given fraction of galaxies within each redshift range with levels varying by 10\%. The dashed line indicates the division line between star-forming and quiescent galaxies proposed by \citet{Ilbert2013}, while solid lines represent our adjusted division lines.
   }
  \label{fig:color}
\end{figure*}

Figure~\ref{fig:color} presents the rest-frame $r-J$ versus ${\rm NUV}-r$ color-color diagrams in three redshift bins, without correcting for dust extinction. The background contours show the distribution of our sample, with levels enclosing 10\%, 20\%, ..., 90\% of galaxies from the innermost to outermost contours. The dashed line indicates the division proposed by \citet{Ilbert2013} to separate star-forming and quiescent populations:
\begin{equation}
{\rm NUV} - r =
\begin{cases}
3.1, & r - J < 0.7, \\
3\,(r - J) + 1, & r - J \geq 0.7.
\end{cases}
\end{equation}
A clear bimodality is present in all three redshift bins. The locations of the two populations shift toward bluer colors with increasing redshift. Consequently, at $0.5 < z \leq 1$, the \citet{Ilbert2013} division traces this valley well and cleanly separates the two populations; at higher redshift, however, the same division progressively deviates from the valley between the two populations and instead intersects the quiescent population, particularly at $2 < z \leq 3$. 
To account for this evolution, we adopt redshift-dependent adjustments to the division proposed by \citet{Ilbert2013} such that it follows the valley between the two populations. For $1 < z \leq 2$, we empirically flatten the slope of the division line. This adjustment alone is sufficient to isolate the quiescent population. The modified division line is defined as follows:
\begin{equation} 
{\rm NUV} - r = 
\begin{cases} 3.1, & r - J < 0.7, \\ 
2.5\,(r - J) + 1.35, & r - J \geq 0.7. 
\end{cases} 
\end{equation}
For $2 < z \leq 3$, adjusting the slope alone is insufficient to effectively isolate the quiescent population. Thus, in addition to a further slope modification, we introduce a redder turning point. The resulting division line is defined as follows:
\begin{equation}
{\rm NUV} - r =
\begin{cases}
3.1, & r - J < 0.78, \\
2.3,(r - J) + 1.306, & r - J \geq 0.78.
\end{cases}
\end{equation}

These adjusted divisions, shown as solid lines in Figure~\ref{fig:color}, more closely follow the valley of the bimodal distribution. Although the adjustment is determined visually, modest variations in the adopted parameters do not significantly affect our results. We therefore classify galaxies above the division as quiescent galaxies (QGs), and those below as star-forming galaxies (SFGs).

The rest-frame {\it J}-band absolute magnitudes are corrected for dust extinction using $E(B-V)$ to obtain $M_J$. From these, we derive the corresponding dust-corrected apparent magnitudes, $m_J$. The intrinsic rest-frame {\it J}-band surface brightness, corrected for both dust extinction and cosmological dimming, is defined as
\begin{equation}
  \mu_J = m_J + 2.5\,\log \left( 2\pi R_{e,\,J}^2 \right) -  2.5\,\log \left(1+z\right)^3, 
\end{equation}
\noindent
where $R_{e,J}$ is in arcsec and the $-2.5\,\log (1+z)^3$ term is to remove cosmological dimming effect (\citealt{Yu2023}; also see next section).

\section{Parametric description}

We adopt a parametric description of the redshift evolution of luminosity, size, and surface brightness. A full derivation, considering JWST instrument parameters, is available in Section~3.1 of \citet{Yu2023}; here we summarize the formalism. 
We consider a galaxy observed at redshift $z$ and assume that its rest-frame luminosity at given rest-frame ${\it J}$ band, evolves as
\begin{equation}\label{lumi}
  L_J = (1+z)^{\alpha} \, L_{J,\,0},
\end{equation}
where the subscript ``0'' denotes the corresponding quantity at $z=0$, and 
$\alpha$ is an empirical parameter describing the average rest-frame $J$-band luminosity evolution at fixed stellar mass, rather than a direct model of the underlying star formation history (SFH). In terms of rest-frame absolute magnitude, Equation~(\ref{lumi}) becomes
\begin{equation}\label{Mevo}
  M_J = M_{J,0} - 2.5\,\log (1+z)^{\alpha}.
\end{equation}
We further assume that the physical size of the galaxy, characterized by the effective radius $R_{e}$, decreases toward higher redshift following a power law of
\begin{equation}\label{Revo}
  R_{e,J} = (1+z)^{\beta} \, R_{e,J,0}.
\end{equation}
The observed surface brightness at redshift $z$ is defined as
\begin{equation}\label{eq:muobs}
  \mu_{{\rm obs}, J} = m_J + 2.5\,\log \left( 2\pi R_{e,J}^2 \right),
\end{equation}
where $m_J$ is the apparent magnitude. Combining the luminosity and size evolution, the observed surface brightness at $z$ and $z=0$ are related by
\begin{equation}\label{full}
  \mu_{{\rm obs}, J} - 2.5\,\log (1+z)^3
  =
  \mu_{ {\rm obs},J,0} - 2.5\,\log (1+z)^{\alpha - 2\beta},
\end{equation}
where $2.5\,\log (1+z)^3$ is the term accounting for cosmological surface-brightness dimming. 
The intrinsic surface brightness at $z$, correcting for cosmological surface-brightness dimming, is therefore defined as
\begin{equation} \label{int}
  \mu_J \equiv 
  \mu_{{\rm obs}, J} - 2.5\,\log (1+z)^3.
\end{equation}
Equation~(\ref{int}) corresponds to the definition of surface brightness adopted in the previous section. 
Combining Equations~(\ref{full}) and (\ref{int}), the galaxy intrinsic surface brightness evolves as
\begin{equation}\label{eq:mu}
  \mu_J = \mu_{J,0} - 2.5\,\log (1+z)^{\gamma},
\end{equation}
where
\begin{equation}\label{eq:con}
  \gamma = \alpha - 2\beta.
\end{equation}

\section{Results} \label{sec:results}

\subsection{Evolution of intrinsic surface brightness}

\begin{figure}
  \centering
  \includegraphics[width=1\columnwidth]{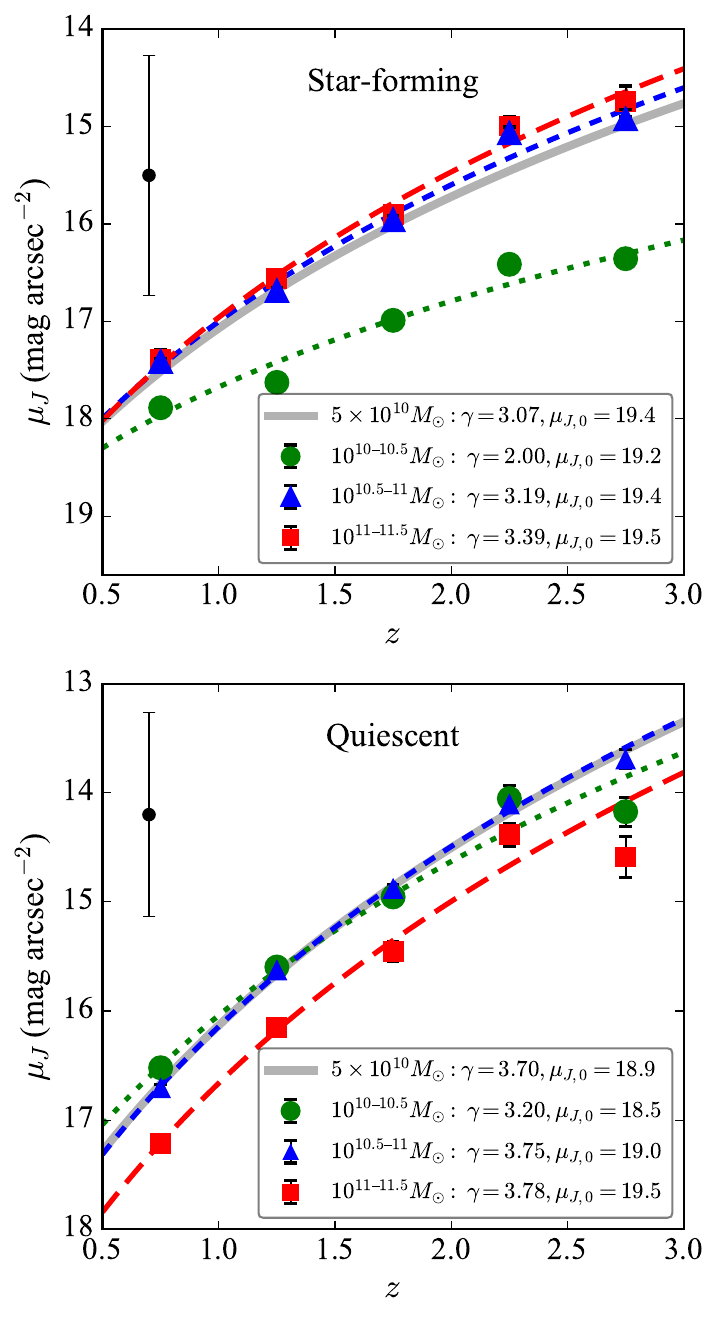}
  \caption{Redshift evolution of intrinsic rest-frame {\it J}-band surface brightness, $\mu_J$, corrected for dust extinction and cosmological dimming.  In each redshift bin, symbols denote the median $\mu_J$, shown as green circles, blue triangles, and red squares from lowest to highest mass bin. 
  The black error bar in the upper-left corner indicates the average $1\sigma$ scatter of the distributions across the redshift bins. The error bars on the data points represent the uncertainties of the medians, which are smaller than or comparable to the symbol size.  The best-fit relation from Equation~(\ref{eq:mu}) for each mass bin is plotted, with parameters given in the legend. 
   }
  \label{fig:mu}
\end{figure}

To investigate the evolution of the intrinsic rest-frame {\it J}-band $\mu_J$, we divide the sample into three stellar mass bins, $10^{10\text{--}10.5}M_\odot$, $10^{10.5\text{--}11}M_\odot$, and $10^{11\text{--}11.5}M_\odot$, and further subdivide each mass bin into five redshift bins uniformly spanning $0.5 \leq z \leq 3$. For each bin, we compute the median $\mu_J$ and estimate its uncertainty as the standard deviation of the distribution divided by the square root of the number of galaxies in that bin. Figure~\ref{fig:mu} presents the redshift dependence of $\mu_J$ for each mass bin, with SFGs and QGs shown in the top and bottom panels, respectively. For both populations, $\mu_J$ systematically becomes brighter with increasing redshift. In particular, SFGs in the two higher-mass bins ($>10^{10.5}\,M_\odot$) exhibit nearly identical median $\mu_J$ values at a fixed redshift and follow consistent evolutionary trends. To quantify this evolution, we fit Equation~(\ref{eq:mu}) to the median values in each mass bin, using the uncertainty as inverse-variance weights, and overplot the resulting best-fit relations as dashed curves in Figure~\ref{fig:mu}; the corresponding parameters are listed in the figure legend and Table~\ref{tbl}.

For SFGs, the best-fit relations for the $10^{10.5\text{--}11}M_\odot$ and $10^{11\text{--}11.5}M_\odot$ bins are highly consistent, with $\gamma=3.19\pm0.08$ and $3.39\pm0.17$, and $\mu_{J,0}=19.4\pm0.07$ and $19.5\pm0.18$, respectively. This similarity indicates that SFGs in the higher-mass range ($10^{10.5\text{--}11.5}M_\odot$) share comparable intrinsic surface brightness and evolutionary behavior up to $z=3$. In contrast, SFGs in the lower-mass bin $10^{10\text{--}10.5}M_\odot$ are systematically fainter, particularly at higher redshift, and exhibit a significantly shallower evolution, with $\gamma=2.00\pm0.05$. The $z=0$ surface brightness, $\mu_{J,0}=19.2\pm0.05$, remains consistent with that of higher-mass systems within uncertainties, indicating that the primary difference arises from the weaker redshift evolution rather than an offset at $z=0$.

For QGs, the two higher-mass bins exhibit similar evolutionary slopes, with $\gamma=3.75\pm0.07$ and $3.78\pm0.13$ for the $10^{10.5\text{--}11}M_\odot$ and $10^{11\text{--}11.5}M_\odot$ bins, respectively. Unlike the case for SFGs, however, the $z=0$ surface brightness of these two bins differs by $\sim0.7$ mag. In contrast, QGs in the lower-mass bin $10^{10\text{--}10.5}M_\odot$ exhibit a shallower evolution, with $\gamma=3.20\pm0.11$ and $\mu_{J,0}=18.5\pm0.09$.
To facilitate comparison with previous studies, in each redshift bin we estimate the median $\mu_J$ at $M_\star = 5\times10^{10}M_\odot$ by interpolating among the median $\mu_J$ values across the mass bins as a function of their median masses.
We then fit Equation~(\ref{eq:mu}) to the resulting $\mu_J$-$z$ relation at $M_\star=5\times10^{10}M_\odot$. For SFGs, the best-fit evolution trend is
\begin{equation}
\mu_J = (19.4\pm0.07) - 2.5\,\log(1+z)^{3.07\pm0.08},
\end{equation}
whereas for QGs, the best-fit relation is
\begin{equation}
\mu_J = (18.9\pm0.07) - 2.5\,\log(1+z)^{3.70\pm0.08}.
\end{equation}
The best-fit parameters for each mass bin, as well as those evaluated at $M_\star=5\times10^{10}M_\odot$, are listed in Table~\ref{tbl}.

Our results are consistent with previous studies showing that galaxy intrinsic surface brightness becomes brighter toward higher redshift \citep[e.g.,][]{Schade1995, Schade1996, Lilly1998, Roche1998, Labbe2003, Barden2005, Ichikawa2010, Sobral2013, Whitney2020}. Using the same functional form as Equation~(\ref{eq:mu}), \citet{Ichikawa2010} reported a much shallower evolution, with $\gamma=1.21\text{--}1.38$ in the rest-frame {\it z} band. This difference likely reflects differences in the measurement approach, as their analysis is based on ground-based Subaru imaging with lower spatial resolution and adopts the Kron radius to estimate surface brightness at rest-frame {\it z} band. The Kron radius can systematically overestimate the measured galaxy size due to PSF effects, particularly at higher redshift, and this consequently leads to a shallower inferred evolutionary trend.

\subsection{Evolution of effective radius}
\label{sec:radius_evolution} 

To isolate the physical drivers governing the evolution of $\mu_J$, it is essential to first quantify the redshift evolution of the rest-frame $J$-band effective radius ($R_{e,J}$). The left column of Figure~\ref{fig:MRe} illustrates the redshift evolution of the median $R_{e,J}$ for SFGs (top panel) and QGs (bottom panel), with errorbars denoting the 1$\sigma$ scatter of the distribution within each redshift bin. The best-fit evolutionary tracks, parameterized by Equation~(\ref{Revo}), are overlaid as dashed curves, with the corresponding best-fit parameters detailed in the legend and Table~\ref{tbl}. 

For SFGs, the redshift evolution of $R_{e,J}$ exhibits a clear dependence on stellar mass. In the higher-mass regimes, the size evolution is relatively steep, scaling as $R_{e,J} \propto (1+z)^{-1.09\pm0.09}$ for $10^{11\text{--}11.5}\,M_\odot$ and $R_{e,J} \propto (1+z)^{-0.95\pm0.04}$ for $10^{10.5\text{--}11}\,M_\odot$; these power-law indices are consistent within the uncertainties. In contrast, galaxies in the lowest-mass bin ($10^{10\text{--}10.5}\,M_\odot$) undergo a significantly shallower evolution, with $R_{e,J} \propto (1+z)^{-0.66\pm0.02}$. This mass-dependent trend, whereby lower-mass SFGs evolve more slowly in size, is in agreement with both legacy HST studies \citep{vanderWel2014} and recent JWST investigations \citep{Martorano2024}. To facilitate a direct comparison with the broader literature, we extract the median effective radius at a fiducial stellar mass of $M_\star = 5\times10^{10}\,M_\odot$, yielding a best-fit redshift evolution for SFGs:
\begin{equation}
  R_{e, J} = (6.0\pm0.18)\,(1+z)^{-0.92\pm0.04}\,{\rm kpc}.
\end{equation}

\begin{figure}
  \centering
  \includegraphics[width=1\columnwidth]{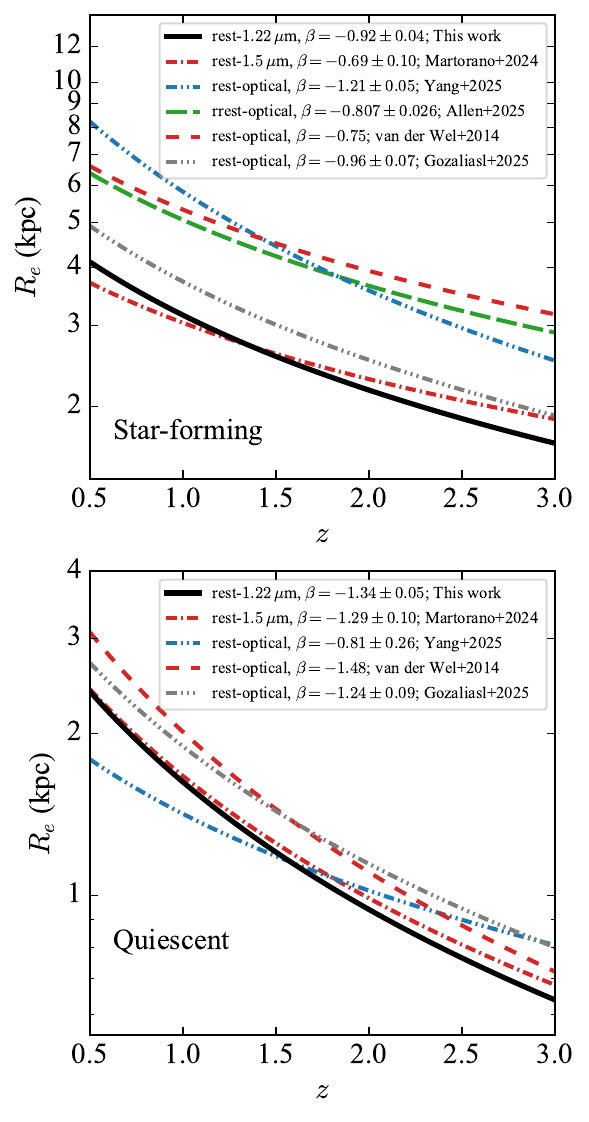}
  \caption{Comparison of the best-fit size evolution at a fixed stellar mass of $M_\star = 5 \times 10^{10} M_\odot$ between this work and previous studies, including \cite{vanderWel2014}, \cite{Martorano2024} (at similar mass but not exactly), \cite{Allen2025}, \cite{Yang2025}, and \cite{Gozaliasl2025}. The results for star-forming and quiescent galaxies are shown in the top and bottom panels, respectively. It noted that the results by \cite{Gozaliasl2025} are for brightest group galaxies.
   }
  \label{fig:Re_comp}
\end{figure}

Our derived evolutionary slope at this fiducial mass is steeper than the rest-frame optical size evolution ($\beta = -0.75$) reported by HST study of \cite{vanderWel2014}, as well as several recent JWST studies \citep{Ward2024, Martorano2024, Allen2025}, which find $\beta$ in the range of $-0.63$ to $-0.807$. In contrast, \citet{Yang2025} report a steeper rest-frame optical evolution of $\beta = -1.21\pm0.05$ and \cite{Gozaliasl2025} reported $-0.96\pm0.07$, which are closer to our measurement. The comparison for SFGs is shown in the top panel of Figure~\ref{fig:Re_comp}.

For QGs, the size evolution is generally steeper than that of SFGs and exhibits only a weak dependence on stellar mass. The two higher-mass bins ($10^{11\text{--}11.5}\,M_\odot$ and $10^{10.5\text{--}11}\,M_\odot$) display identical slopes of $R_{e,J} \propto (1+z)^{-1.32\pm0.09}$ and $(1+z)^{-1.38\pm0.04}$, respectively, while the lowest-mass bin ($10^{10\text{--}10.5}\,M_\odot$) follows a modestly shallower trend of $(1+z)^{-1.03\pm0.06}$. The reduced mass dependence relative to SFGs implies that the structural evolution of QGs proceeds more uniformly across the stellar mass. Evaluating the size evolution at the fiducial mass of $M_\star = 5\times10^{10}\,M_\odot$, we obtain the best-fit relation for QGs:
\begin{equation}
  R_{e, J} = (4.1\pm0.16)\,(1+z)^{-1.34\pm0.05}\,{\rm kpc},
\end{equation}
which indicates rapid size growth toward lower redshifts. 

The derived evolutionary slope for QGs, $\beta = -1.34\pm0.05$, is in good agreement with rest-frame optical HST constraints ($\beta = -1.48$; \citealt{vanderWel2014}). Furthermore, our measurement is consistent with recent JWST findings at rest-frame $1.5$ $\mu$m, which report $\beta = -1.29\pm0.10$ \citep{Martorano2024} and that for brightest QGs at rest-optical, which report $\beta=-1.24\pm0.09$ \citep{Gozaliasl2025}.
In contrast, \citet{Yang2025} find a significantly shallower evolution ($\beta = -0.81\pm0.26$); this discrepancy is likely driven by their large uncertainty, due to the limited statistics based on the small sample of QGs at $z > 2$. The comparison for QGs is shown in the bottom panel of Figure~\ref{fig:Re_comp}.

Galaxy color gradients, with redder stellar populations typically more centrally concentrated, imply that galaxies appear smaller at longer wavelengths \citep[e.g.,][]{Kelvin2012,Wuyts2012}. Although the rest-frame $J$ band is a better tracer of the underlying stellar mass distribution than rest-frame optical bands, it may still be affected by young stellar populations and residual star-forming light, yielding slightly larger sizes than those measured at longer wavelengths. Based on radiative-transfer calculations of TNG50 galaxies, \citet{Baes2024} showed that rest-frame $J$-band effective radii are only systematically larger than $K_s$-band effective radii by $\sim$7--9\%, depending on color. Thus, the size overestimate caused by using the rest-frame $J$ band should be small and is unlikely to significantly affect the evolutionary slope measured in this work.

\subsection{Evolution of luminosity}

\begin{figure*}
  \centering
  \includegraphics[width=2\columnwidth]{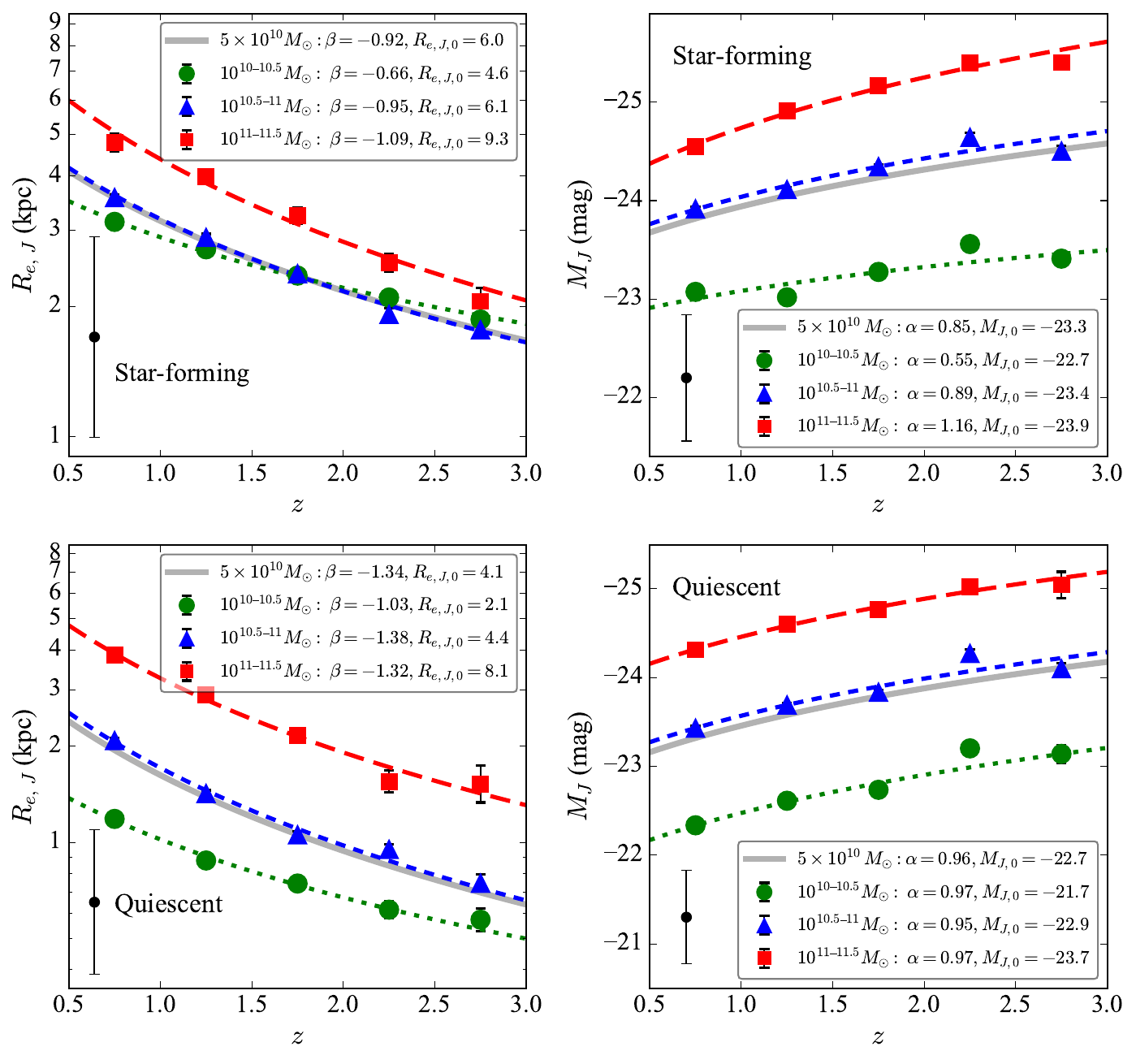}
  \caption{Redshift evolution of rest-frame {\it J}-band galaxy effective radius ($R_{e, J}$; left) and dust-extinction-corrected absolute magnitude ($M_J$; right).   In each redshift bin, symbols denote the median $R_{e,J}$ or $M_J$, with green circles, blue triangles, and red squares corresponding to increasing stellar mass. The black error bar in the bottom-left corner indicates the average $1\sigma$ scatter of the distributions across the redshift bins. The error bars on the data points represent the uncertainties of the medians, which are smaller than or comparable to the symbol size.  The best-fit relations from Equation~(\ref{Revo}) for $R_{e,J}$ and Equation~(\ref{Mevo}) for $M_J$ are overplotted with the corresponding parameters listed in the legend.
   }
  \label{fig:MRe}
\end{figure*}

The redshift evolution of the dust-extinction-corrected rest-frame {\it J}-band luminosity $L_J$ is mathematically equivalent to that of the absolute magnitude $M_J$. The evolution of $M_J$ is shown in the right column of Figure~\ref{fig:MRe}, with symbols indicating medians in stellar mass and redshift bins, and errorbars representing the $1\sigma$ scatter. The corresponding best-fit relations, described by Equation~(\ref{Mevo}), are overplotted as dashed curves. For both SFGs (top panel) and QGs (bottom panel), galaxies of a given stellar mass become systematically brighter toward higher redshift. The slope of this evolution depends on stellar mass, with more massive systems exhibiting steeper trends. For SFGs, the evolution slope decreases from $\alpha=1.16\pm0.07$ in the highest mass bin ($10^{11\text{--}11.5}M_\odot$) to $\alpha=0.89\pm0.04$ at $10^{10.5\text{--}11}M_\odot$, and further to $\alpha=0.55\pm0.03$ at $10^{10\text{--}10.5}M_\odot$. QGs show a similar but overall shallower dependence on stellar mass, with $\alpha=0.97\pm0.08$, $0.95\pm0.04$, and $0.97\pm0.06$ across the same mass bins, respectively.

To facilitate a direct comparison independent of mass binning, we also derive the best-fit evolution at a fixed stellar mass of $M_\star=5\times10^{10}M_\odot$. For SFGs, the evolution is described by
\begin{equation}
  M_J = (-23.3\pm0.04) - 2.5\,\log(1+z)^{0.85\pm0.04},
\end{equation}
while QGs follow 
\begin{equation}
  M_J = (-22.7\pm0.04) - 2.5\,\log(1+z)^{0.96\pm0.04},
\end{equation}
These results indicate comparable evolutionary slopes for the two populations at fixed mass, despite their offset in $z=0$ magnitude. 
Our results are consistent with \cite{Sachdeva2013}, who show that the absolute magnitude gets brighter at higher redshift based on HST imaging. 
We note that the derived slopes are systematically higher by $\sim0.3$--$0.4$ than those reported by \citet{Yu2023}, based on the 3D-HST catalog \citep{Brammer2012, Skelton2014}. This difference is likely driven by the limited wavelength coverage of HST imaging, which does not probe the rest-frame near-infrared at $z\sim1$--3, and thus may introduce biases in tracing the underlying stellar mass distribution at cosmic noon.

\renewcommand{\arraystretch}{1.1}
\begin{deluxetable*}{ccccccccc}
\tablecaption{
Best-fit parameters of the evolutionary relations (Equations~(\ref{Mevo}), (\ref{Revo}), and (\ref{eq:mu})) for each mass bin and at $M_\star=5\times10^{10}M_\odot$.
\label{tbl}}
\tablehead{
\colhead{Mass} & \colhead{Type} & \colhead{$\gamma$} & \colhead{$\mu_{J,0}$}&  \colhead{$\alpha$} & \colhead{$M_{J,0}$}& \colhead{$\beta$}&  \colhead{$R_{e,J,0}$} & \colhead{$\alpha - 2\beta$}\\
\colhead{($M_\odot$)} & \colhead{} & \colhead{} & \colhead{(mag arcsec$^{-2}$)} & \colhead{} & \colhead{(mag)} & \colhead{} & \colhead{(kpc)} & \colhead{} \\
\colhead{(1)} & \colhead{(2)} & \colhead{(3)} & \colhead{(4)} & \colhead{(5)}& \colhead{(6)}& \colhead{(7)}& \colhead{(8)} & \colhead{(9)}
}
\startdata
$10^{10\text{--}10.5}$ & SFGs & $2.00\pm0.05$ & $19.2\pm0.05$ & $0.55\pm0.03$ & $-22.7\pm0.03$ & $-0.66\pm0.02$ & $4.6\pm0.10$ & $1.87\pm0.05$ \\
$10^{10.5\text{--}11}$ & SFGs & $3.19\pm0.08$ & $19.4\pm0.07$ & $0.89\pm0.04$ & $-23.4\pm0.04$ & $-0.95\pm0.04$ & $6.1\pm0.20$ & $2.79\pm0.09$ \\
$10^{11\text{--}11.5}$ & SFGs & $3.39\pm0.17$ & $19.5\pm0.18$ & $1.16\pm0.07$ & $-23.9\pm0.07$ & $-1.09\pm0.09$ & $9.3\pm0.75$ & $3.34\pm0.19$ \\
$5\times10^{10}$ & SFGs & $3.07\pm0.08$ & $19.4\pm0.07$ & $0.85\pm0.04$ & $-23.3\pm0.04$ & $-0.92\pm0.04$ & $6.0\pm0.18$ & $2.69\pm0.09$ \\
\hline 
$10^{10\text{--}10.5}$ & QGs & $3.20\pm0.11$ & $18.5\pm0.09$ & $0.97\pm0.06$ & $-21.7\pm0.05$ & $-1.03\pm0.06$ & $2.1\pm0.10$ & $3.03\pm0.13$ \\
$10^{10.5\text{--}11}$ & QGs & $3.75\pm0.07$ & $19.0\pm0.07$ & $0.95\pm0.04$ & $-22.9\pm0.04$ & $-1.38\pm0.04$ & $4.4\pm0.17$ & $3.71\pm0.09$ \\
$10^{11\text{--}11.5}$  & QGs & $3.78\pm0.13$ & $19.5\pm0.12$ & $0.97\pm0.08$ & $-23.7\pm0.07$ & $-1.32\pm0.09$ & $8.1\pm0.58$ & $3.61\pm0.20$ \\
$5\times10^{10}$ & QGs & $3.70\pm0.08$ & $18.9\pm0.07$ & $0.96\pm0.04$ & $-22.7\pm0.04$ & $-1.34\pm0.05$ & $4.1\pm0.16$ & $3.64\pm0.11$ \\
\enddata
\end{deluxetable*}

\section{Discussion}

\subsection{Implication from the size evolution}

Tracing the redshift evolution of the galaxy size distribution may impose critical constraints on disk assembly history. In the standard hierarchical framework, disks for a given circular velocity forming at earlier epochs are intrinsically more compact, as the initial disk scale length scales inversely with the Hubble parameter at the formation epoch \citep{Mo1998}. However, while this foundational model establishes an accurate baseline for initial disk sizes based on angular momentum conservation, it does not track subsequent structural evolution via gas accretion, mergers, or radial migration. Therefore, to fully understand the compact sizes of high-redshift disks with their present-day counterparts, additional evolutionary mechanisms must be invoked \citep{Lilly2016}.

A potential driver of this subsequent size evolution is inside-out growth, wherein late-accreting, high-angular-momentum gas from the circumgalactic medium preferentially settles in the disk outskirts \citep[e.g.,][]{Guo2011, Tacchella2015, Nelson2016}. Recent JWST observations have provided direct validation of this scenario. For instance, \citet{Baker2025} reported a strongly rising specific star-formation rate (sSFR) gradient at $z = 7.43$, marking the first direct detection of inside-out growth during the epoch of reionization, while \citet{Parlanti2025} and \cite{Xiao2025}  identified inside-out stellar age gradients in spiral galaxies at $z = 2.2$ and $z=5.2$, respectively. Nevertheless, the evolutionary picture remains complex. At $z \geq 2$, mildly negative sSFR gradients have also been observed, suggesting that pure in-situ inside-out growth cannot universally account for size evolution beyond cosmic noon \citep{Song2025}. 

Radial migration provides an additional pathway for disk expansion. Secular processes driven by bars and spiral structure can redistribute stellar mass through corotation scattering \citep{Debattista2006, Roskar2008, Minchev2012}. Although such processes can be efficient in evolved disks, their impact at high redshift may be limited if non-axisymmetric structures are not yet fully developed. In contrast, violent disk instabilities offer a rapid mechanism at early times. Massive clumps can efficiently exchange angular momentum with the disk, driving inward clump migration and outward stellar redistribution \citep{Bournaud2007, Bournaud2010, Bournaud2011, Bournaud2016, Wu2020}. Furthermore, a sustained increase in star formation rate driven by instabilities in clumpy galaxies may also contribute to galaxy growth \citep{Faisst2025}. Recent analyses of radial profile of surface brightness and color index suggest a dual-channel picture: at cosmic noon, clump-driven migration dominates size growth in roughly 60\% of star-forming disks, while secular processes already operate in the remaining systems \citep{XuYu2024, Yu2025}.

In addition, mergers, mostly minor mergers, are also responsible for the size growth, particularly in massive quiescent galaxies \citep{Faisst2017, Carollo2013, Carollo2016, Fagioli2016}. In this scenario, the accretion of low-mass satellites deposits stars at large galactocentric radii, leading to an efficient increase in effective radius together with the growth of mass.

These considerations collectively predict smaller galaxy sizes at higher redshift. Consistent with this expectation, we find that at $M_\star = 5 \times 10^{10}M_\odot$, star-forming galaxies follow $R_{e,J} \propto (1+z)^\beta$ with $\beta = -0.92 \pm 0.04$. This slope is consistent with early studies based on rest-frame UV sizes of Lyman break galaxies at $z\sim2\text{--}6$, which yield $\beta \approx -1.1$ \citep{Giavalisco1996, Ferguson2004, Oesch2010, Mosleh2012}, and is comparable to the rest-optical evolution reported for ultra-massive systems, $\beta = -1.18 \pm 0.15$ at $M_\star > 10^{11.4}M_\odot$ \citep{Faisst2017}. Figure~\ref{fig:Re_comp} compares the size evolution at $M_\star = 5 \times 10^{10}M_\odot$ derived in this work with previous studies.
 Our derived slope is steeper than the rest-frame optical relation at $M_\star = 5 \times 10^{10}M_\odot$ from HST $\beta=-0.75$ \citep{vanderWel2014} and also steeper than several recent JWST measurements, including $\beta=-0.63\pm0.07$ \citep{Ward2024} and $\beta=-0.69\pm0.1$ at rest-frame 1.5 $\mu$m \citep{Martorano2024}. Interestingly, \citet{Martorano2024} also report a steeper evolution ($\beta=-1.16\pm0.31$) for the higher-mass bin ($10^{11\text{--}11.5} M_\odot$), consistent with our result ($\beta=-1.09\pm0.09$) in the same mass range. 
 Our derived $\beta = -0.92$ is slightly higher than the rest-optical evolution slope reported by \citet{Allen2025} ($\beta=-0.807\pm0.026$). In contrast, based on the COSMOS-Web data as probed in this study, \cite{Yang2025} derived higher value $\beta=-1.21\pm0.05$ for SFGs, while, focusing on brightest SFGs in galaxy groups, \cite{Gozaliasl2025} obtained $\beta=-0.96\pm0.07$, fully consistent with our results.

 The size evolution of galaxies has also been investigated in cosmological simulations \citep{Costantin2023, LaChance2025, Shen2024}. In particular, \citet{Costantin2023} analyzed the IllustrisTNG TNG50-1 simulation, the highest-resolution realization within the IllustrisTNG framework, focusing on star-forming galaxies with $M_\star = 10^{9.5}\text{--}10^{12}M_\odot$ over $3 \leq z \leq 6$. They reported an evolution slope of $\beta = -1.15$ in the F356W filter, corresponding approximately to rest-frame optical wavelengths. This value is only modestly steeper than our measured value, suggesting that current high-resolution simulations broadly reproduce the observed size growth of star-forming galaxies.

 Our result therefore lies between existing JWST estimates and helps narrow the current dispersion in reported slopes. We suggest that this intermediate value is primarily driven by our use of rest-frame {\it J}-band sizes. At $\sim1.2$ \micron, these measurements are less sensitive to spatial variations in dust attenuation and star formation, and thus better trace the underlying stellar mass distribution than rest-optical sizes.

Furthermore, lower-mass systems ($10^{10}$--$10^{10.5}M_\odot$) exhibit significantly slower size evolution ($\beta=-0.66\pm0.02$) compared to their higher-mass counterparts ($\beta\approx-1.0$). 
The slower size evolution of low-mass star-forming galaxies relative to their high-mass counterparts arises from a confluence of mass-dependent physical processes. Supernova feedback drives large-scale gas outflows in shallow dark matter potential wells, suppressing the cold gas supply needed for outside-in disk growth \citep[e.g.,][]{Wyithe2013, Vogelsberger2013}. Furthermore, low-mass galaxies are dominated by in-situ star formation fed by centrally concentrated cold-mode accretion \citep{Keres2009, Kocjan2024}, and their merger rates are too low to contribute the dry minor mergers that efficiently grow the effective radii of more massive systems \citep{Bedorf2013, Conselice2014}.

In contrast, quenched galaxies exhibit a steeper and more uniform evolution, with $\beta = -1.34 \pm 0.05$. As shown in the bottom panel of Figure~\ref{fig:Re_comp}, this is in good agreement with HST-based measurements in the rest-frame optical ($\beta=-1.48$; \citealt{vanderWel2014}) and consistent with recent JWST results at rest-frame 1.5 $\micron$ ($\beta=-1.29\pm0.10$; \citealt{Martorano2024}) and rest-optical wavelength ($\beta=-1.24\pm0.09$; \citealt{Gozaliasl2025}).
The consistency across datasets and wavelengths indicates that the structural evolution of quenched galaxies is both strong and well regulated since $z\sim3$, with limited sensitivity to measurement band.

We note that inside-out growth, radial migration, clump-driven evolution, and merger-driven ex-situ growth can all affect galaxy sizes, and their signatures may be degenerate in global structural quantities such as $R_e$, $M_J$, and $\mu_J$. Quantifying the contribution of each process therefore requires more direct diagnostics, including spatially resolved stellar population gradients, low-surface-brightness outskirts and tidal features, and resolved stellar kinematics. For example, ultra-deep imaging can constrain stellar haloes and outer envelopes associated with merger-driven growth \citep{Buitrago2017}, while simulation-based analyses can probe the origin of the mass-size relation and its scatter \citep[e.g.,][]{Du2024}, and trace the in-situ and ex-situ origin of stellar mass and quantify the effect of mergers on galaxy structure \citep{Angeloudi2025}. Such analyses are beyond the scope of the present work, but are needed to quantify the relative roles of secular processes, clump-driven evolution, and mergers in the observed size evolution.

\subsection{Effect of Bulges}

The sizes in this work are measured from single-Sérsic fitting, a convenient and widely used parameterization for large galaxy samples \citep[e.g.,][]{vanderWel2014,Martorano2024,Yang2025}. However, a single Sérsic model does not separate bulge, disk, and bar components, which may have distinct star formation histories and evolutionary paths \citep[e.g.,][]{Kormendy2012,GonzalezDelgado2014}. Therefore, the measured effective radius should be interpreted as a global half-light radius rather than the size of a specific structural component. For SFGs, which are generally disk dominated, the single-Sérsic radius is expected to mainly trace the disk light, although it can still be affected by central bulges and bars. Thus, the inferred size evolution of SFGs reflects the combined effects of disk growth and bulge assembly, but is expected to be primarily driven by disk growth. As a first test, we performed a bulge-disk decomposition using an exponential disk and a de Vaucouleurs bulge. At $M_\star=5\times10^{10}M_{\odot}$, we find disk-size evolution $R_{e,J,{\rm disk}} \propto (1+z)^{-0.89\pm0.03}$, broadly consistent with the single-Sérsic result. This supports disk growth as the dominant driver of SFG size evolution. Future work will use more sophisticated multi-component decompositions, allowing flexible bulge Sérsic indices and including bar components where appropriate, since neglecting bars can bias bulge-disk decompositions \citep{Mendez-Abreu2008,GaoHo2017}.

\subsection{The Drivers of Surface Brightness Evolution}

Our analysis reveals a strong brightening of the intrinsic rest-frame $J$-band surface brightness toward higher redshifts, which we parameterize as $\mu_J \propto -2.5 \log(1+z)^\gamma$. We derive steep evolutionary slopes of $\gamma = 3.07 \pm 0.08$ for SFGs and $\gamma = 3.70 \pm 0.08$ for QGs. Analytically, the evolution of surface brightness ($\gamma$), effective radius ($\beta$), and luminosity ($\alpha$) are explicitly coupled via the relation $\gamma = \alpha - 2\beta$ (see final column of Table~\ref{tbl}). By isolating the luminosity evolution, which scales as $L_J \propto (1+z)^\alpha$ with $\alpha = 0.85 \pm 0.04$ for SFGs and $\alpha = 0.96 \pm 0.04$ for QGs, we demonstrate that while galaxies were intrinsically more luminous at early epochs, the dramatic brightening of $\mu_J$ is predominantly driven by the strong size evolution discussed in the previous section, with luminosity enhancement providing a secondary contribution.

Our derived evolutionary slopes $\alpha$ are systematically higher by $\sim 0.3$--$0.4$ than those reported by \citet{Yu2023} based on the 3D-HST catalog \citep{Brammer2012, Skelton2014}, likely reflecting the use of pre-JWST measurements in that study. \citet{Yu2023} further show that the evolution is more rapid in bluer wavebands. A consistent physical picture emerges in which both results are driven by the redshift evolution of the star-forming main sequence \citep{Speagle2014, Popesso2023, Scoville2023}: at higher redshift, galaxies exhibit elevated star formation rates at fixed stellar mass, leading to enhanced luminosity even in the rest-frame {\it J} band. This interpretation is supported by recent literature demonstrating that the star formation rate (SFR) surface density increases significantly at earlier epochs \citep{Calabro2024, Yang2025}. Furthermore, \citet{Yang2025} showed that more massive galaxies exhibit a steeper redshift evolution in SFR surface density. This is consistent our findings, as we observe a correspondingly steeper luminosity evolution slope $\alpha$ for galaxies in the more massive bins, bridging the gap between structural compaction and enhanced star formation efficiency.
We note that converting $\alpha$ into a specific SFH would require additional assumptions about stellar population evolution, dust attenuation, and mass-to-light ratio variations. Therefore, $\alpha$ should be interpreted as an empirical constraint on luminosity growth that galaxy evolution models should reproduce jointly with the observed evolution of stellar mass and size. Establishing a direct connection between $\alpha$ and detailed SFHs will require future forward modeling.

A particularly striking feature of our analysis is the behavior of the most massive SFGs ($M_\star > 10^{10.5}\ M_{\odot}$). At any fixed redshift, these massive systems exhibit very similar $\mu_J$ values, showing little dependence on stellar mass. This reveals a tightly regulated structural assembly process at the high-mass end. For massive SFGs, size expansion and luminosity growth operate largely in lockstep; the two evolutionary vectors nearly balance each other, maintaining an almost mass-independent intrinsic surface brightness at a given epoch.

\section{Summary} \label{sec:summary}

Galaxy size and surface brightness are fundamental observables that can be directly measured from imaging data, and their redshift evolution provides key constraints on galaxy formation and evolution. In this work, we analyze a sample of 15,420 galaxies with stellar masses of $10^{10}$--$10^{11.5}\ M_{\odot}$ over $0.5 \leq z \leq 3$ from the JWST COSMOS-Web survey. We adopt effective radius measurements from \citet{Yu2026} in the NIRCam bands and interpolate them to the rest-frame {\it J} band to obtain $R_{e,J}$. The intrinsic surface brightness $\mu_J$ is corrected for both dust extinction and cosmological dimming.

We divide the sample into three stellar mass bins ($10^{10\text{--}10.5}M_\odot$, $10^{10.5\text{--}11}M_\odot$, and $10^{11\text{--}11.5}M_\odot$) and examine the redshift evolution of $R_{e,J}$, $\mu_J$, and the rest-frame {\it J}-band luminosity. Our main results are summarized as follows:

\begin{enumerate}
  \item At a characteristic mass of $M_\star = 5 \times 10^{10}\ M_{\odot}$, star-forming galaxies follow $R_{e,J} \propto (1+z)^\beta$ with $\beta = -0.92 \pm 0.04$. This slope is steeper than recent rest-frame near-infrared JWST measurements \citep{Martorano2024}, but shallower than rest-frame optical results \citep{Yang2025}. The differences likely reflect systematic effects among studies, including variations in wavelength, sample selection, and stellar mass estimates. Quiescent galaxies exhibit more rapid size evolution, with $\beta = -1.34 \pm 0.05$, consistent with previous work.
  
  \item Among star-forming galaxies, lower-mass systems ($10^{10}$--$10^{10.5}\ M_{\odot}$) show significantly slower size evolution ($\beta = -0.66 \pm 0.02$) compared to higher-mass galaxies ($\beta \approx -1$). In contrast, quiescent galaxies evolve more rapidly across all mass bins, with slopes that are systematically steeper by $\sim0.3$--0.5 than those of star-forming galaxies.
  
  \item At $M_\star = 5 \times 10^{10}\ M_{\odot}$, the rest-frame {\it J}-band luminosity increases with higher redshift following $L_J \propto (1+z)^\alpha$, with $\alpha = 0.85 \pm 0.04$ for star-forming galaxies and $\alpha = 0.96 \pm 0.04$ for quiescent galaxies. This trend likely reflects elevated star formation rate surface densities at higher redshifts \citep{Yang2025}. 
  
  \item At $M_\star = 5 \times 10^{10}\ M_{\odot}$, the intrinsic surface brightness increases toward higher redshift, following $\mu_J \propto -2.5 \log(1+z)^\gamma$. We obtain $\gamma = 3.07 \pm 0.08$ for star-forming galaxies and $\gamma = 3.70 \pm 0.08$ for quiescent galaxies. Massive star-forming galaxies ($M_\star > 10^{10.5}\ M_{\odot}$) exhibit nearly identical $\mu_J$ at fixed redshift, independent of mass. We further find that $\gamma \approx \alpha - 2\beta$, as expected from the definition of surface brightness, implying that its evolution is governed by the combined evolution of galaxy size and luminosity. 
\end{enumerate}

Overall, by utilizing rest-frame {\it J}-band measurements to approximately trace the underlying stellar mass distribution, we provide new constraints on the structural and photometric history of galaxies out to $z=3$. Our results highlight a pronounced size evolution across both star-forming and quiescent populations, demonstrating that the observed scaling of surface brightness is a direct manifestation of this structural growth coupled with luminosity evolution. We note that analysis at fixed stellar mass do not follow the evolution of individual galaxy populations; future studies using approaches such as constant comoving number density matching or dark matter halo matching are needed to directly connect progenitors and descendants across cosmic time.

\begin{acknowledgments}
We thank the referee for the insightful comments and suggestions.
This work is supported by the National SKA Program of China No. 2025SKA0150103, National Natural Science Foundation of China under Nos. 12550002, 12133008, 12221003, 11890692. We acknowledge the science research grants from the China Manned Space Project with No. CMS-CSST-2021-A04 and No. CMS-CSST-2025-A10.
SYU acknowledges the support from the start-up funding provided by Xiamen University. LM acknowledges the financial contribution from the PRIN-MUR 2022 20227RNLY3 grant ``The concordance cosmological model: stress-tests with galaxy clusters'' supported by Next Generation EU and from the grant ASI n. 2024-10-HH.0 ``Attività scientifiche per la missione Euclid – fase E''. This project has received funding from the European Union’s Horizon 2020 research and innovation programme under the Marie Skłodowska-Curie grant agreement No 101148925. The JWST data presented in this article were obtained from the Mikulski Archive for Space Telescopes (MAST) at the Space Telescope Science Institute. The specific observations analyzed can be accessed via \dataset[DOI: 10.17909/ph8h-qf05]{https://doi.org/10.17909/ph8h-qf05}. 
\end{acknowledgments}

\begin{contribution}
SYU conceived the project, performed the data reduction and analysis, and wrote the manuscript. All authors contributed to the interpretation of the results and to the manuscript.

\end{contribution}

\facilities{JWST(NIRCam) }

\software{astropy \citep{Astropy2013,Astropy2018,Astropy2022},  
          IMFIT \citep{Erwin2015}, 
          SEP \citep{Bertin1996, Barbary2016} }

\appendix

\setcounter{figure}{0}
\renewcommand{\thefigure}{A\arabic{figure}}

\section{Additional plots}

In this appendix, we illustrate the wavelength dependence of the measured effective radius. We select galaxies at $1<z<1.5$ to balance two requirements. This interval is narrow enough to reduce contamination from the intrinsic redshift evolution of galaxy size, which would otherwise be mixed with the wavelength dependence of $R_e$ in a broader redshift range. At the same time, the four NIRCam filters still span a broad rest-frame wavelength range, from the optical to approximately the $K_s$ band, allowing us to illustrate the dependence of $R_e$ on $\lambda_{\rm rest}$. For each galaxy, we use the best-fit $R_e$ measured in the F115W, F150W, F277W, and F444W bands and assign each measurement to its corresponding rest-frame wavelength, defined as the central wavelength of the filter divided by $(1+z)$. We then combine the measurements from all galaxies and all filters, separately for SFGs and QGs, and divide them into four bins of rest-frame wavelength. In each bin, we calculate the median $R_e$ and the upper and lower $1\sigma$ scatter of the distribution.

\begin{figure}
  \centering
  \includegraphics[width=1\columnwidth]{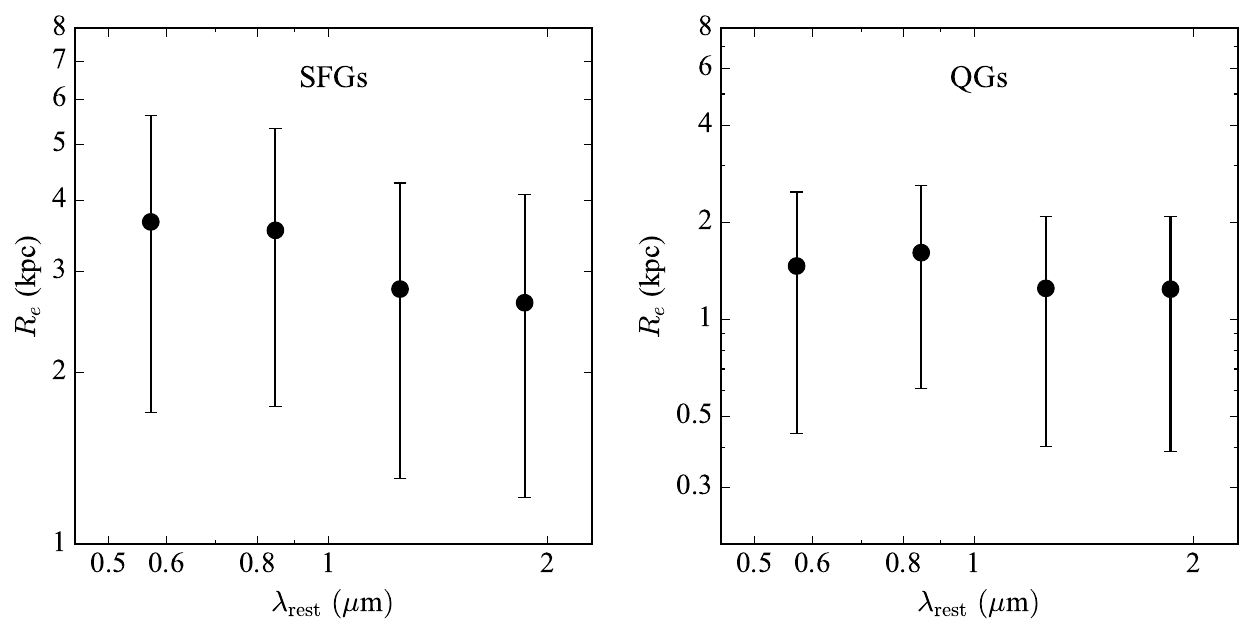}
  \caption{Average dependence of $R_e$ on rest-frame wavelength for SFGs (left) and QGs (right) at $1<z<1.5$. The data points show the median $R_e$ in each rest-frame wavelength bin. The upper and lower error bars indicate the upper and lower $1\sigma$ scatter of the distribution in each bin.}
  \label{fig:Re_lambda}
\end{figure}

Figure~\ref{fig:Re_lambda} presents the results for SFGs in the left panel and QGs in the right panel. For SFGs, $R_e$ decreases toward longer rest-frame wavelengths, likely reflecting radial color gradients associated with inside-out galaxy formation \citep[e.g.,][]{Guo2011,Tacchella2015,Nelson2016}. For QGs, the wavelength dependence is much weaker, consistent with weaker radial color gradients.  We note, however, that the dependence of $R_e$ on wavelength is not perfectly described by a linear relation between $\log R_e$ and $\log \lambda_{\rm rest}$. This may partly reflect chromatic surface-brightness modulation, in which spatial variations in stellar populations, dust attenuation, and nebular emission modify the observed surface-brightness profile and hence the measured structural parameters in a filter-dependent manner \citep{Papaderos2023}. Thus, while our rest-frame $J$-band interpolation reduces morphology $k$-corrections by measuring sizes at a fixed rest-frame wavelength, it does not fully remove this effect. We therefore regard chromatic surface-brightness modulation as a potential source of systematic uncertainty in the inferred $R_{e,J}$ values and their redshift evolution.

\end{document}